\documentclass[twocolumn,showpacs,preprintnumbers,amsmath,amssymb,prb]{revtex4}

\usepackage{graphicx}
\usepackage{dcolumn}
\usepackage{bm}
\usepackage{tabularx}
\usepackage{amsmath}

\begin{document}

\title{Full superconducting gap in the doped topological crystalline insulator, Sn$_{0.6}$In$_{0.4}$Te}

\author{L. P. He, Z. Zhang, J. Pan, X. C. Hong, S. Y. Zhou, and S. Y. Li$^*$}

\affiliation{State Key Laboratory of Surface Physics, Department of
Physics, and Laboratory of Advanced Materials, Fudan University,
Shanghai 200433, P. R. China}

\date{\today}

\begin{abstract}
The thermal conductivity of the doped topological crystalline
insulator, Sn$_{0.6}$In$_{0.4}$Te superconducting single crystal
with $T_c =$ 4.1 K, was measured down to 50 mK. It is found that the
residual linear term $\kappa_0/T$ is negligible in zero magnetic
field. The $\kappa_0/T$ shows a slow field dependence at low
magnetic field. These results suggest that the superconducting gap
is nodeless, unless there exist point nodes with directions
perpendicular to the heat current. Due to its high-symmetry fcc
crystal structure of Sn$_{0.6}$In$_{0.4}$Te, however, such point
nodes can be excluded. Therefore we demonstrate that this
topological superconductor candidate has a full superconducting gap
in the bulk. It is likely the unconventional odd-parity $A_{1u}$
state which supports a surface Andreev bound state.
\end{abstract}

\pacs{74.25.fc, 74.62.Dh}

\maketitle

\section{Introduction}

Three dimensional (3D) topological insulators (TIs) have attracted
great attentions in recent years because of their novel quantum
states. \cite{MZHasan,SCZhang} They are characterized by an inverted
gap in the bulk caused by strong spin-orbit coupling (SOC) and a
gapless surface state protected by time-reversal symmetry.
\cite{MZHasan,SCZhang} In the layered compound Bi$_2$Se$_3$, a
typical 3D TI, superconductivity was found by intercalating Cu atoms
between the Se layers. \cite{YSHor} Cu$_x$Bi$_2$Se$_3$ is considered
as a promising candidate of topological superconductor,
\cite{YSHor,MKriener,PDas,SSasaki} with potential applications in
topological quantum computing. \cite{JAlicea,CWJBeenakker} Recently,
Fu proposed a new type of materials named topological crystalline
insulator (TCI), in which the gapless surface state is protected by
mirror symmetry of the crystal, instead of time-reversal symmetry.
\cite{LFu} Later, SnTe with face-centered-cubic (fcc) crystal
structure was predicted to be such a TCI,\cite{THHsieh} and this
prediction was soon confirmed by angle-resolved photoemission
spectroscopy (ARPES) measurements. \cite{YTanaka}

Superconductivity exists in Sn$_{1-x}$In$_x$Te, with Sn partially
substituted by In.
\cite{Bushmarina,ASErickson,GBalakrishnan,RDZhong} The
superconducting transition temperature $T_c$ is about 1.2 K at low
doping $x =$ 0.045. \cite{SSasaki2} With increasing the doping
level, $T_c$ increases to a maximum, about 4.6 K at $x =$ 0.45.
\cite{RDZhong} Zero-bias conductance peak (ZBCP) was observed in
Sn$_{1-x}$In$_x$Te ($x =$ 0.045) by using point-contact spectroscopy
technique which suggests the existence of Andreev bound state (ABS).
\cite{SSasaki2} The ARPES measurements also support the presence of
topological surface state in Sn$_{1-x}$In$_x$Te ($x =$ 0.045) by
comparing to pristine SnTe. \cite{TSato} These results indicate that
Sn$_{1-x}$In$_x$Te is another candidate of topological
superconductor. \cite{SSasaki2,TSato}

Although the surface state of Sn$_{1-x}$In$_x$Te has been clarified,
the superconducting gap in its bulk is still unknown. Theoretically,
an unconventional odd-parity $A_{1u}$ state with full
superconducting gap in the bulk is usually required for topological
superconductors. \cite{XLQi,L.Fu} However, the odd-parity pairing
state with point nodes in the gap ($A_{2u}$ state) is also allowed.
\cite{SSasaki,SSasaki2} In this context, it is very important to
determine the superconducting gap structure of Sn$_{1-x}$In$_x$Te in
the bulk.

Ultra-low-temperature thermal conductivity measurement is a bulk
technique to probe the gap structure of superconductors.
\cite{Shakeripour} A negligible residual linear term $\kappa_0/T$ in
zero magnetic field is a strong evidence for nodeless
superconducting gap. Line nodes will result in a finite and
universal (impurity independent) $\kappa_0/T$. \cite{Shakeripour}
Point nodes will cause a nonuniversal finite $\kappa_0/T$ if the
impurity scattering rate is high, unless the direction of point
nodes is perpendicular to the heat current. \cite{Shakeripour}
Furthermore, the field dependence of $\kappa_0/T$ can give more
information on nodal gap, the gap anisotropy, or multiple gaps.
\cite{Shakeripour}

In this paper, we present the ultra-low-temperature thermal
conductivity measurements of Sn$_{0.6}$In$_{0.4}$Te single crystal
with $T_c =$ 4.1 K, which has a relatively low residual resistivity
thus suits for heat transport study. We find a negligible
$\kappa_0/T$ in zero field and a slow field dependence of
$\kappa_0/T$ at low field. By considering its high-symmetry fcc
crystal structure, we exclude superconducting gap with point nodes
($A_{2u}$ state), and conclude that Sn$_{0.6}$In$_{0.4}$Te has a
full superconducting gap in the bulk ($A_{1u}$ state).

\section{Experiment}

The Sn$_{0.6}$In$_{0.4}$Te single crystals were grown by modified
Bridgman method. \cite{GBalakrishnan} The dc magnetic susceptibility
was measured by using a superconducting quantum interference device
(MPMS, Quantum Design). The (100) plane was identified by X-ray
diffraction measurements. The sample for transport measurements was
cut and polished to a rectangular shape of 2.6 $\times$ 1.6 mm$^2$
in the (100) plane, with the thickness of 0.31 mm. Four silver wires
were attached to the sample with silver paint, which were used for
both resistivity and thermal conductivity measurements, with
electrical and heat currents in the (100) plane. The contacts are
metallic with typical resistance 20 m$\Omega$ at 2 K. The thermal
conductivity was measured in a dilution refrigerator, using a
standard four-wire steady-state method with two RuO$_2$ chip
thermometers, calibrated {\it in situ} against a reference RuO$_2$
thermometer. Magnetic fields were applied perpendicular to the heat
current. To ensure a homogeneous field distribution in the sample,
all fields for resistivity and thermal conductivity measurements
were applied at temperature above $T_c$.

\section{Results and Discussion}

\begin{figure}
\includegraphics[clip,width=7.5cm]{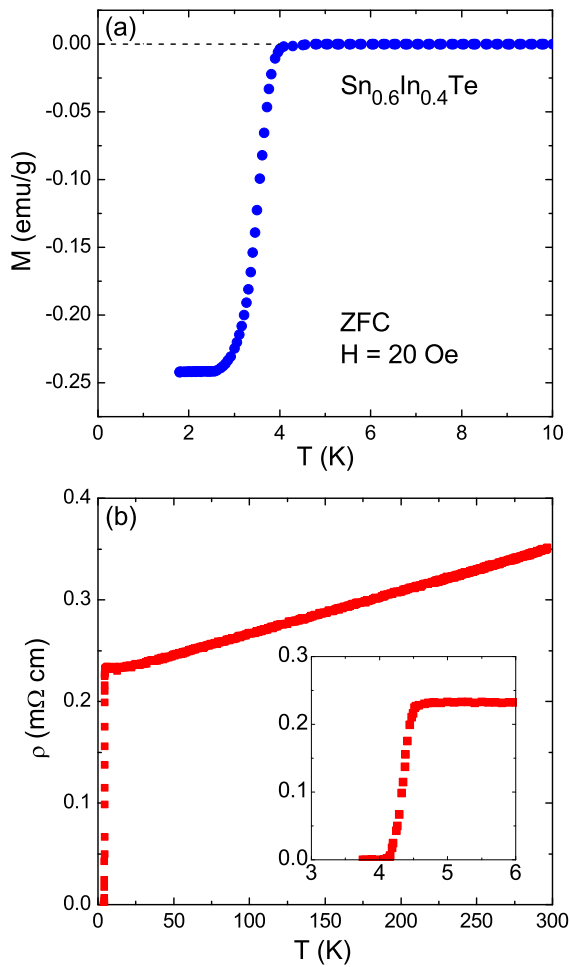}
\caption{(Color online). (a) Low-temperature magnetization of
Sn$_{0.6}$In$_{0.4}$Te single crystal measured with
zero-field-cooled (ZFC) process. (b) Temperature dependence of
resistivity of Sn$_{0.6}$In$_{0.4}$Te single crystal at zero field.
Insert shows the resistive superconducting transition at low
temperature.}
\end{figure}

\begin{figure}
\includegraphics[clip,width=7.27cm]{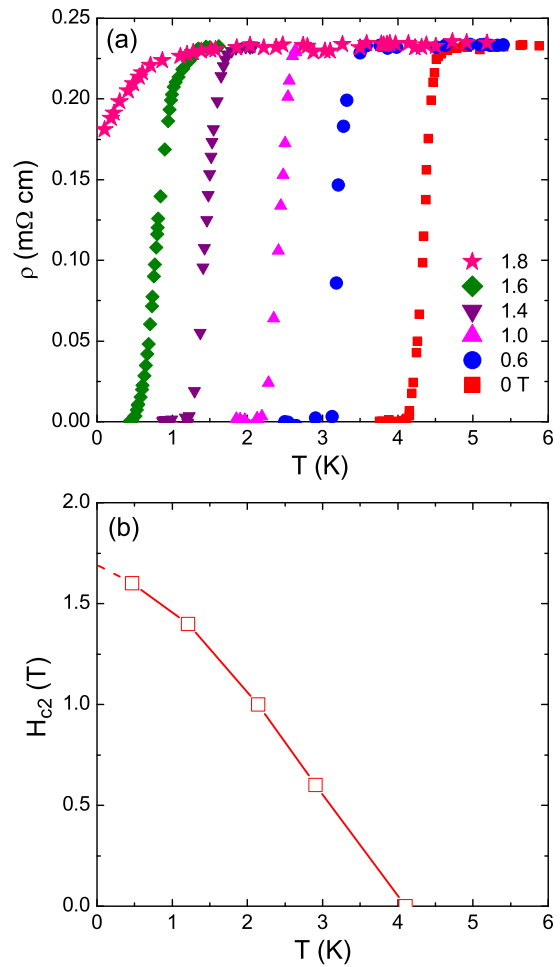}
\caption{(Color online). (a) Low-temperature resistivity of
Sn$_{0.6}$In$_{0.4}$Te single crystal in magnetic field up to 1.8 T. (b)
Temperature dependence of the upper critical field $H_{c2}(T)$,
defined by $\rho = 0$. The dashed line is a guide to the eye, which
points to $H_{c2}(0) \approx$ 1.7 T.}
\end{figure}

Figure 1(a) shows the low-temperature magnetization of
Sn$_{0.6}$In$_{0.4}$Te single crystal. The onset of superconducting
transition is at 4.0 K. The magnetization saturates below 2.6 K and
the shielding fraction at 2 K is estimated to be $\sim 93\%$,
showing the bulk superconductivity in our sample. The temperature
dependence of resistivity at zero field is shown in Fig. 1(b). From
the inset of Fig. 1(b), the width of the resistive superconductivity
transition (10-90$\%$) is 0.27 K, and the $T_c$ defined by $\rho =
0$ is 4.1 K. The $\rho(T)$ curve from $T_c$ to 10 K is quite flat,
which extrapolates to a residual resistivity $\rho_0$ = 0.233
m$\Omega$ cm. Thus the residual resistivity ratio $\rho$(300
K)/$\rho_0\approx$ 1.5 is obtained. This value is slightly larger
than that in Ref. 14, indicating higher quality of our sample.

Figure 2(a) shows the low-temperature resistivity of
Sn$_{0.6}$In$_{0.4}$Te single crystal in magnetic fields up to 1.8
T. To estimate the upper critical field $H_{c2}$(0), the temperature
dependence of $H_{c2}(T)$, defined by $\rho = 0$, is plotted in Fig.
2(b). $H_{c2}(0) \approx$ 1.7 T is roughly estimated. A slightly
different $H_{c2}(0)$ does not affect our discussion on the field
dependence of $\kappa_0/T$ below.

\begin{figure}
\includegraphics[clip,width=8.5cm]{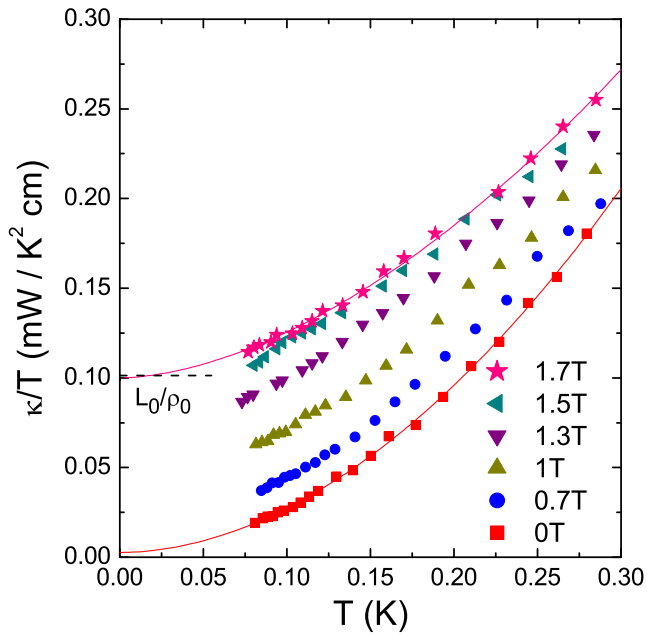}
\caption{(Color online). Low-temperature thermal conductivity of
Sn$_{0.6}$In$_{0.4}$Te single crystal in magnetic fields applied
perpendicular to (100) plane. The two solid lines represent the fit
curves of $H =$ 0 and 1.7 T, respectively. The dashed lines are the
normal-state Wiedemann-Franz law expectation $L_0$/$\rho_0$, with
the Lorenz number $L_0 =$ 2.45 $\times$ 10$^{-8}$W $\Omega$ K$^{-2}$
and $\rho_0$ = 0.233 m$\Omega$ cm.}
\end{figure}

The temperature dependence of thermal conductivity for
Sn$_{0.6}$In$_{0.4}$Te single crystal in zero and magnetic fields
are plotted as $\kappa/T$ vs $T$ in Fig. 3. The measured thermal
conductivity contains two contributions, $\kappa$ = $\kappa_e$ +
$\kappa_p$, which come from electrons and phonons, respectively. In
order to separate the two contributions, all the curves below 0.3 K
were fitted to $\kappa/T$ = $a +
bT^{\alpha-1}$.\cite{MSutherland,SYLi} The two terms $aT$ and
$bT^{\alpha}$ represent contributions from electrons and phonons,
respectively. The residual linear term $\kappa_0/T \equiv a$ is
obtained by extrapolated $\kappa/T$ to $T$ = 0 K. Because of the
specular reflections of phonons at the sample surfaces, the power
$\alpha$ in the second term is typically between 2 and 3.
\cite{MSutherland,SYLi}

\begin{figure}
\includegraphics[clip,width=8.3cm]{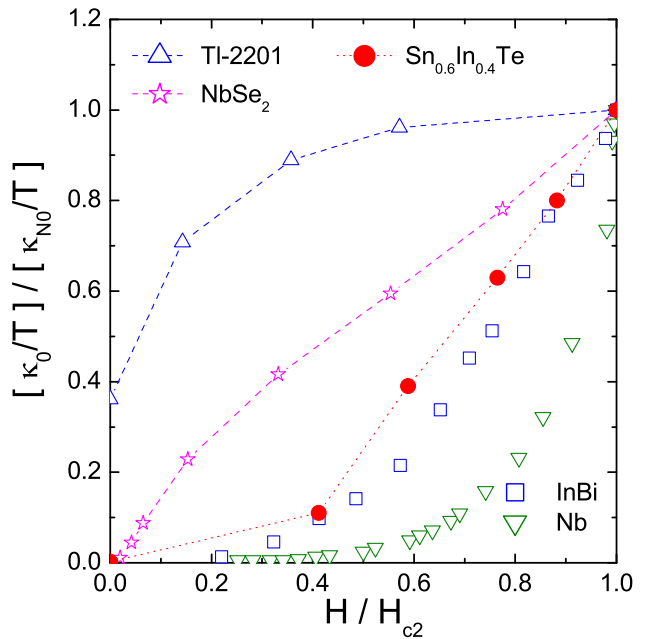}
\caption{(Color online). Normalized residual linear term
$\kappa_0/T$ of Sn$_{0.6}$In$_{0.4}$Te as a function of $H/H_{c2}$.
For comparison, similar data are shown for the clean $s$-wave
superconductor Nb, \cite{Lowell} the dirty $s$-wave superconducting
alloy InBi, \cite{JOWillis} the multiband $s$-wave superconductor
NbSe$_2$, \cite{EBoaknin} and an overdoped $d$-wave cuprate
superconductor Tl-2201. \cite{Proust}} \end{figure}

We first check the high-field normal-state thermal conductivity. In
$H_{c2}(0) =$ 1.7 T, the fitting gives $\kappa_0/T =$ 0.100 $\pm$
0.004 mW K$^{-2}$ cm$^{-1}$ and $\alpha = 2.45 \pm 0.05$. This value
of $\kappa_0/T$ roughly agrees with the normal-state Wiedemann-Franz
law expectation $L_0$/$\rho_0 =$ 0.105 mW K$^{-2}$ cm$^{-1}$, with
the Lorenz number $L_0 =$ 2.45 $\times$ 10$^{-8}$W $\Omega$ K$^{-2}$
and $\rho_0$ = 0.233 m$\Omega$ cm. The verification of the
Wiedemann-Franz law in the normal state demonstrates that our
thermal conductivity measurements are reliable. For the power
$\alpha$, previously $\alpha \approx$ 2.2 was found in the $s$-wave
superconductor Cu$_x$TiSe$_2$, \cite{S.Y.Li} and recently, $\alpha
\approx$ 2 has been observed in some iron-based superconductors such
as BaFe$_{1.9}$Ni$_{0.1}$As$_2$, \cite{LDing} KFe$_2$As$_2$,
\cite{JKDong} and Ba(Fe$_{1-x}$Ru$_x$)$_2$As$_2$. \cite{XQiu} Below
we will only concentrate on electron contribution.

In zero field, $\kappa_0/T$ = 0.003 $\pm$ 0.002 mW K$^{-2}$
cm$^{-1}$ is obtained by the fitting. Note that this value is within
our experimental error bar $\pm$ 0.005 mW K$^{-2}$ cm$^{-1}$.
Therefore the $\kappa_0/T$ of Sn$_{0.6}$In$_{0.4}$Te in zero field
is negligible, comparing to the normal-state $\kappa_0/T$ = 0.100 mW
K$^{-2}$ cm$^{-1}$ in $H_{c2}(0) =$ 1.7 T. For $s$-wave nodeless
superconductors, there are no fermionic quasiparticles to conduct
heat when $T \rightarrow 0$ since all electrons become Cooper pairs.
Therefore, there is no residual linear term of $\kappa_0/T$, as seen
in Nb and InBi. \cite{Lowell,JOWillis} In contrast, a finite
$\kappa_0/T$ in zero field is usually observed in a superconductor
with nodal gap, if the heat current is not perpendicular to the
nodal directions. \cite{Shakeripour} For example, $\kappa_0/T$ =
1.41 mW K$^{-2}$ cm$^{-1}$ was observed with heat current in the
$ab$ plane for the overdoped cuprate Tl$_2$Ba$_2$CuO$_{6+\delta}$
(Tl-2201), a $d$-wave superconductor with $T_c$ = 15 K.
\cite{Proust} However, if the heat current is perpendicular to the
nodal directions, the $\kappa_0/T$ will still be zero despite the
existence of gap nodes.\cite{Shakeripour} The negligible
$\kappa_0/T$ of our Sn$_{0.6}$In$_{0.4}$Te single crystal in zero
field suggests a nodeless superconducting gap, but the existence of
nodes with directions perpendicular to the heat current is still
possible.

Since both unconventional odd-parity states $A_{1u}$ (full
superconducting gap) and $A_{2u}$ (gap with point nodes) are allowed
in Sn$_{1-x}$In$_x$Te from previous surface
experiments,\cite{SSasaki2,TSato} here we show how the $A_{2u}$
state can be excluded from our bulk measurements by further
considering its impurity scattering rate and symmetry of crystal
structure. For a superconductor with point nodes in the gap, the
observation of a finite $\kappa_0/T$ requires a relatively high
impurity scattering rate. \cite{MJGraf} We estimate the normal state
scattering rate $\Gamma_0$ of our Sn$_{0.6}$In$_{0.4}$Te from
$\rho_0$ and the plasma frequency $\omega_p$ = $c$/$\lambda_0$, in
which $c$ is the speed of light and $\lambda_0$ is the penetration
depth when $T \rightarrow 0$ ($\lambda_0 =$ 860 nm according to ref.
14). With the formula $\Gamma_0 =$ $(\omega_p^2/8\pi)\rho_0$,
$\hbar$$\Gamma_0$/$k_B$$T_c$ = 2.9 is obtained. This value is high
enough to lead to a finite $\kappa_0/T$.\cite{XGLuo,MATanatar} Since
we do not observe $\kappa_0/T$ in Sn$_{0.6}$In$_{0.4}$Te, the
directions of the point nodes, if they do exist, must be
perpendicular to the heat current. However, due to the high-symmetry
fcc crystal structure of Sn$_{1-x}$In$_x$Te, such a configuration is
impossible. In other words, one can not let all the directions of
point nodes be perpendicular to the heat current at the same time.
For example, in our heat transport experiments, if there is point
nodes along the [100] direction which is perpendicular to our heat
current, there must also be point nodes along the [010] and [001]
directions, which can not be perpendicular to our heat current
simultaneously. From above analysis, we can safely exclude the
$A_{2u}$ state with point nodes in Sn$_{0.6}$In$_{0.4}$Te, and its
superconducting state should be the fully gapped $A_{1u}$ state.

The field dependence of $\kappa_0/T$ will give more information of
the superconducting gap structure. Between $H$ = 0 and 1.7 T, we fit
all the curves and obtain the $\kappa_0/T$ for each magnetic field.
The normalized $\kappa_0(H)/T$ as a function of $H/H_{c2}$ for
Sn$_{0.6}$In$_{0.4}$Te is plotted in Fig. 4. For comparison, similar
data of the clean $s$-wave superconductor Nb, \cite{Lowell} the
dirty s-wave superconducting alloy InBi, \cite{JOWillis} the
multiband s-wave superconductor NbSe$_2$, \cite{EBoaknin} and an
overdoped $d$-wave cuprate superconductor Tl-2201, \cite{Proust} are
also plotted. The slow field dependence of $\kappa_0/T$ at low field
for Nb and InBi manifests their full superconducting
gap.\cite{Lowell,JOWillis} From Fig. 4, the curve of the normalized
$\kappa_0(H)/T$ for Sn$_{0.6}$In$_{0.4}$Te is close to that of the
dirty $s$-wave superconductor InBi, which further supports a full
superconducting gap.

To check whether Sn$_{0.6}$In$_{0.4}$Te is a dirty superconductor,
we estimate its superconducting coherence length $\xi_0$ and the
electron mean free path $l$. From $H_{c2}(0) =$ 1.7 T, we obtain
$\xi_0 \approx$ 142 {\AA} through the relation $H_{c2}(0) =$
$\Phi_0/2\pi\xi_0^2$. The electron mean free path is estimated using
the normal-state thermal conductivity $\kappa_N$, specific heat $c$,
and the Fermi velocity $v_F$. Since $\kappa_N =$ ($1/3$)$cv_Fl$, we
have the relation $l =$ 3($\kappa_N/T$)$/$($\gamma$$v_F$),
\cite{S.Y.Li} where $\gamma =$ $c/T$ is the linear specific heat
coefficient and $v_F$ can be calculated through $\xi_0 \equiv$
$\hbar$$v_F/\pi\Delta(0)$ and $\Delta(0)$ = 1.76$k_B$$T_c$.
\cite{MTinkham} With $\kappa_N/T =$ 0.100 mW K$^{-2}$ cm$^{-1}$,
$\gamma =$ 2.62 mJ mol$^{-1}$ K$^{-2}$, \cite{GBalakrishnan} and the
calculated $v_F =$ 0.28 eV {\AA}, we get $l \approx$ 11 {\AA}. Since
$l \ll \xi_0$, Sn$_{0.6}$In$_{0.4}$Te is indeed a dirty
superconductor.

\section{Summary}

In summary, we have measured the thermal conductivity of
Sn$_{0.6}$In$_{0.4}$Te single crystal down to 50 mK. The negligible
$\kappa_0/T$ at zero field and the slow field dependence of
$\kappa_0/T$ at low field suggest nodeless superconducting gap, and
the possibility of point nodes perpendicular to the heat current is
excluded due to the high-symmetry fcc crystal structure. Combining
with previous surface experiments, our new bulk measurements
demonstrate that the superconducting state in Sn$_{1-x}$In$_x$Te is
likely an unconventional odd-parity $A_{1u}$ state with fully
superconducting gap, if it is indeed a topological superconductor.

\begin{center}
{\bf ACKNOWLEDGEMENTS}
\end{center}
This work is supported by the Natural Science Foundation of China,
the Ministry of Science and Technology of China (National Basic
Research Program No: 2009CB929203 and 2012CB821402), and the Program
for Professor of Special Appointment (Eastern Scholar) at Shanghai
Institutions of Higher Learning. \\

$^*$ E-mail: shiyan$\_$li@fudan.edu.cn

\end{document}